# Bounding the Uncertainty of Graphical Games: The Complexity of Simple Requirements, Pareto and Strong Nash Equilibria

**Gianluigi Greco**
Dept. of Mathematics, University of Calabria
I-87030, Rende, Italy
ggreco@mat.unical.it

**Francesco Scarcello**
DEIS, University of Calabria
I-87030, Rende, Italy
scarcello@deis.unical.it

## Abstract

We investigate the complexity of bounding the uncertainty of graphical games, and we provide new insight into the intrinsic difficulty of computing Nash equilibria. In particular, we show that, if one adds very simple and natural additional requirements to a graphical game, the existence of Nash equilibria is no longer guaranteed, and computing an equilibrium is an intractable problem. Moreover, if stronger equilibrium conditions are required for the game, we get hardness results for the second level of the polynomial hierarchy. Our results offer a clear picture of the complexity of mixed Nash equilibria in graphical games, and answer some open research questions posed by Conitzer and Sandholm (2003).

## 1 Introduction

Several classes of games with succinct representation have been recently proposed in the literature with the aim of modeling games with a large number of players (e.g., Vickrey and Koller 2002, Kearns and Mansour 2002, Leyton-Brown and Tennenholtz 2003, Gal and Pfeffer 2004). In this paper, we deal with the class of *graphical games* (firstly formalized by Koller and Milch 2001, and Kearns, Littman, and Singh 2001), where the utility function of each player $p$ is defined only in terms of those players $p$ is directly interested in. Within this setting, a widely accepted formalization of rational behavior for players is the notion of Nash equilibrium, introduced by Nash (1951). Note that this notion was originally exploited for modeling the outcome of *strategic games* where utility functions are extensively represented through a single table having an entry for *each* combination of players' strategies. In fact, many interesting questions concerning Nash equilibria of graphical games have been lately faced. For instance, Fabrikant, Papadimitriou, and Talwar (2004), Alvarez, Gabarro, and Serna (2005) and Gottlob, Greco, and Scarcello (2003) considered the existence of *pure* Nash equilibria, i.e., equilibria where each player must play in a non-aleatory manner.

The first crucial question is deciding the existence of a Nash equilibrium for a given graphical game. Indeed, although the existence of a Nash equilibrium is always guaranteed for some kind of randomized (*mixed*) strategies by Nash's famous theorem, in many real-world applications the existence of "any" equilibrium is not enough. Rather, it is often useful to single out equilibria satisfying some additional, application-oriented constraints. For instance, a game where players choose their actions just tossing a coin is of no interest in several situations. In these cases, it would be useful to know whether there are other Nash equilibria or, instead, such a fully random equilibrium is the only possible (Nash) outcome of the game at hand.

First results on the computational complexity of equilibria that satisfy some additional requirements have been presented by Gilboa and Zemel (1989) in the context of strategic *two-players* games. Recently, Conitzer and Sandholm (2003) reconsidered such setting and proposed a single reduction (from satisfiability of Boolean formulas) that sharpened most of the results of Gilboa and Zemel (1989) and provided novel ones. It is worthwhile noting that, since the number of players is fixed, the above complexity results strongly rely on the assumption of an unbounded number of actions available for each player. Therefore, they cannot be exploited for studying the complexity of graphical games with many players and a small (possibly fixed) number of actions. Indeed, characterizing the complexity of graphical games has been left as an open research problem by Conitzer and Sandholm (2003).

In fact, the problem of computing special Nash equilibria in graphical games has been originally raised by Kearns, Littman, and Singh (2001), who considered tree-like game structures where each player has just two possible actions, and described a polynomial-time algorithm for computing Nash equilibria, even in presence of some simple kind of constraints. Moreover, the complexity of *constrained* Nash equilibria in graphical games has been recently investigated by Greco and Scarcello (2004) for the setting of pure strategies. However, a complete characterization of the complexity of equilibria with additional requirements was still missing for the case of mixed strategies. For instance, it remained open whether some of the additional constraints introduced by Conitzer and Sandholm (2003), such as constraints on the expected social welfare and on the expected utility for players, are sufficient to make graphical games hard, and whether classical refinements of Nash equilibria, such as Pareto (Aumann 1959) or strong Nash equilibria (Maskin 1985), are any harder.

In this paper, we face these research questions for graphical games. We provide a clear picture of the intrinsic difficulty of bounding the uncertainty of graphical games by asking for additional requirements, either in terms of constraints on actions and payoffs, or in terms of stronger notions of equilibria. In summary, for any graphical game $\mathcal{G}$:

**1)** We show that deciding the existence of a Nash equilibrium satisfying additional requirements is NP-hard, even if there is only one constraint on a single action, and only two available actions for each player.

**2)** We consider constraints on sets of players. In particular, we study the problem of deciding whether, given a set of players $P$ and a Nash equilibrium, there exists another Nash equilibrium for $\mathcal{G}$ where some player in $P$ chooses a different strategy. We prove that this problem, called ANOTHER-NASH, is NP-hard.

**3)** Given a set of players $P$, we may be interested in Nash equilibria where the non-random constraint for $P$ is satisfied, that is, there is at least one player who does not choose her actions in a fully random way. We prove that deciding the existence of such mixed Nash equilibria is NP-hard (problem NON-RANDOM).

**4)** We study Pareto Nash equilibria, i.e., those equilibria such that there exists no other Nash equilibrium where each player gets a strictly higher utility. Note that a game has a Pareto Nash equilibrium if and only if it has a Nash equilibrium, and thus always, in the mixed strategies setting. Yet, we prove that checking whether a profile is a Pareto equilibrium is co-NP-hard.

**5)** The latter result may be extended to strong Nash equilibria, i.e., those profiles where no change of strategies of whatever coalition $C$ of players can simultaneously increase the utility for all players in $C$. However in this case the existence of an equilibrium does not follow from Nash's theorem, and in fact we have here an additional orthogonal source of complexity, making this problem hard for the second level of the polynomial hierarchy. In particular, we show that deciding whether $\mathcal{G}$ has a strong Nash equilibrium is $\Sigma_2^P$-hard.

Before detailing the above contributions, we next formalize our framework and, specifically, the kinds of constraints we shall consider for graphical games.

## 2 Graphical Games

### 2.1 Nash Equilibria

A *graphical game* $\mathcal{G}$ is a tuple $\langle P, Neigh, Act, U \rangle$, where $P$ is a non-empty set of distinct players, $Neigh$ and $Act$ are functions, and $U$ is a set of functions. In particular, for each player $i \in P$: $Neigh$ provides a set of players $Neigh(i) \subseteq P - \{i\}$, called neighbors of $i$; $Act(i)$ defines the set of her possible actions; and $U$ contains her utility function $u_i : Act(i) \times_{j \in Neigh(i)} Act(j) \to \Re$.

The interaction among players of $\mathcal{G}$ is usually represented by an undirected graph $G(\mathcal{G}) = (P, E)$, called a *dependency graph*, whose vertices coincide with the players of $\mathcal{G}$, and $\{p, q\} \in E$ if $p$ is a neighbor of $q$, i.e., $p \in Neigh(q)$.

Let $\mathcal{G} = \langle P, Neigh, Act, U \rangle$ be a game. Each player $i$ may choose an action $a \in Act(i)$ with a given probability $p_a$, where $0 \leq p_a \leq 1$. An individual *strategy* for $i$ is any set $S$ such that: for each $a \in Act(i)$, $S$ contains exactly one pair $(p_a, a)$, and $\sum_{(p_a, a) \in S} p_a = 1$. The set of all the (individual) strategies for $i$ is denoted by $St(i)$. For a non-empty set of players $P' \subseteq P$, a *combined strategy* (also, *profile*) for $P'$ is a set containing exactly one strategy for each player in $P'$. $St(P')$ denotes the set of all combined strategies for the players in $P'$. The combined strategy $\mathbf{x}$ is called *global* if $P' = P$. A global strategy $\mathbf{x}$ such that the individual strategy of each player $i$ contains a pair $(1, a)$ for some action $a$, is called a *pure strategy*, and we say that $i$ *plays* $a$ (in $\mathbf{x}$). Let $i$ by a player, $u_i$ the utility function of $i$, and $\mathbf{x}$ a combined strategy for a set of players $P' \supseteq Neigh(i) \cup \{i\}$. Given an action $a$ in $Act(i)$, we denote by $i_a(\mathbf{x})$ (or even simply by $i_a$, if $\mathbf{x}$ is clear from the context) the probability that player $i$ plays $a$ in the strategy $\mathbf{x}$.

The *payoff* of player $i$ w.r.t. $\mathbf{x}$, denoted by $pay_i(\mathbf{x})$, is the expected value of her utility function given the probability distribution of the actions played by her neighbors in $Neigh(i)$ and by herself, provided by their individual strategies in **x**, in $St(Neigh(i) \cup \{i\})$, i.e., $pay_i(\mathbf{x}) = \mathbf{E}_\mathbf{x}[u_i]$. Note that, if **x** is a pure combined strategy, the payoff $pay_i(\mathbf{x})$ of a player $i$ w.r.t. **x** simply coincides with the value assumed by the utility function $u_i$ on the restriction of **x** to its domain, that is, on the restriction to the actions played by $i$ and by her neighbors in $Neigh(i)$.

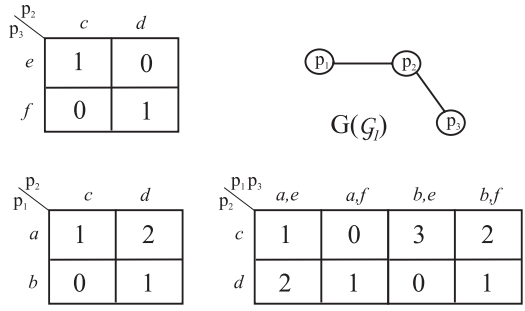

Figure 1: Utilities and dependency graph for $\mathcal{G}_1$.

**Example 2.1** Consider the simple game $\mathcal{G}_1$ among players $p_1$, $p_2$ and $p_3$, with $Neigh(p_1) = \{p_2\}$, $Neigh(p_2) = \{p_1, p_3\}$, and $Neigh(p_3) = \{p_2\}$. Assume that $Act(p_1) = \{a,b\}$, $Act(p_2) = \{c,d\}$, $Act(p_3) = \{e,f\}$, and assume that the utility functions for the players are those shown in the tables reported in Figure 1. Each table shows the utilities for the player whose actions are reported in the rows, while the strategies for her neighbors vary on the columns. Note that, even though $p_1$ is directly interested only in $p_2$, her strategy is transitively affected by $p_3$ as well.   □

Let us now formally define the main concept of equilibrium studied in this paper. Let **x** be a global strategy, $i$ a player, and **y** an individual strategy for $i$. Then, we denote by $\mathbf{x}_{-i}[\mathbf{y}]$ the global strategy where the individual strategy of player $i$ in **x** is replaced by **y**.

**Definition 2.2** Let $\mathcal{G} = \langle P, Neigh, A, U \rangle$ be a game and **x** be a global strategy for $\mathcal{G}$. Then, **x** is a Nash equilibrium for $\mathcal{G}$ if, $\forall i \in P$, $\nexists \mathbf{y} \in St(i)$ such that $pay_i(\mathbf{x}) < pay_i(\mathbf{x}_{-i}[\mathbf{y}])$.   □

Thus, we have a Nash equilibrium if we are in a state where no player has an incentive to deviate from her current choice. For instance, the strategy in which $p_1$, $p_2$ and $p_3$ play $a$, $d$, and $f$, respectively, is an equilibrium for the game $\mathcal{G}_1$.

## 2.2 Constraints, Pareto and Strong Nash Equilibria

Let $\mathcal{G} = \langle P, Neigh, Act, U \rangle$ be a game and $P'$ be a non-empty subset of the players. An *evaluation function* $f_{P'}$ for players in $P'$ is any polynomial-time computable function that, for each combined strategy **x** for the players in $\bigcup_{p \in P'} Neigh(p) \cup P'$, maps the set $\{pay_i(\mathbf{x}) \mid i \in P'\}$ to a real number.

A *constraint on the payoffs* of the players in $P'$ is an expression $c$ of the form $[f_{P'} \text{ op } k]$, with $k \in \Re$ and $\text{op} \in \{<,>,=,\neq,\leq,\geq\}$. The semantics is as follows: a Nash equilibrium **x** satisfies $c$, denoted by $\mathbf{x} \models c$, if $f_{P'}(\mathbf{x}) \text{ op } k$. For instance, if $op(c)$ is $<$, we require that the evaluation of $f_{P'}$ on the Nash equilibrium **x** is less than $k$. Moreover, a *constraint on the actions* of a player $i$ is an expression $c$ of the form $[p_a(i) \text{ op } k]$, with $k \in \Re$ and $\text{op} \in \{<,>,=,\neq,\leq,\geq\}$. A Nash equilibrium **x** satisfies $c$, denoted by $\mathbf{x} \models c$, if $i$ plays $a \in Act(i)$ in **x** with a probability $p_a(i)$ such that $p_a(i) \text{ op } k$. Intuitively, $c$ can be used for looking for equilibria satisfying a given condition on the probability of choosing an action. For instance, a constraint of the form $p_a(i) = 0$ means that we are looking for Nash equilibria in which action $a$ cannot be played by $i$, whereas a constraint of the form $p_a(i) = 1$ means that $i$ is forced to play $a$. The constraint $[p_a(i) \neq \frac{1}{|Act(i)|}]$, for some action $a \in Act(i)$, is what we call the *non-random constraint* for player $i$. Indeed, it means that $i$ cannot play in a fully random way, and has to give more chances to some of her possible actions.

Given a game $\mathcal{G}$ and a set of constraints $\mathcal{C}$, a *constrained Nash equilibrium* is any Nash equilibrium for $\mathcal{G}$ that satisfies all the constraints in $\mathcal{C}$.

For some kind of additional requirements, the original notion of Nash equilibrium is not sufficient to model the game properly, and there is a need for stronger refinements, like Pareto and strong Nash equilibria. In these cases, we ask for equilibria **x** that cannot assign better payoffs to players, even if either all of them, or any coalition of players are allowed to deviate from the profile **x** (see Gottlob, Greco, and Scarcello 2003, for a recent detailed study of these equilibria in the framework of pure strategies). Formally, any set of players $K \subseteq P$ is a *coalition*. Let **x** be a global strategy, $K$ a coalition, and **y** a combined strategy for $K$. Then, we denote by $\mathbf{x}_{-K}[\mathbf{y}]$ the global strategy where, for each player $p \in K$, her individual strategy $p_a \in \mathbf{x}$ is replaced by her individual strategy $p_b \in \mathbf{y}$. If $K$ is a singleton $\{p\}$, we will simply write $\mathbf{x}_{-p}[\mathbf{y}]$.

**Definition 2.3** Let $\mathcal{G} = \langle P, Neigh, A, U \rangle$ be a game and **x** be a global strategy for $\mathcal{G}$. Then,
- **x** is a strong Nash Equilibrium for $\mathcal{G}$ if, $\forall K \subseteq P$, $\forall y \in St(K)$, $\exists p \in K$ such that $pay_p(\mathbf{x}) \geq pay_p(\mathbf{x}_{-K}[y])$ or, equivalently, if $\forall K \subseteq P$, $\nexists y \in St(K)$ such that, $\forall p \in K$, $pay_p(\mathbf{x}) < pay_p(\mathbf{x}_{-K}[y])$.
- **x** is a Pareto Nash Equilibrium for $\mathcal{G}$ if there does not exist a Nash equilibrium **y** for $\mathcal{G}$ such that, $\forall p \in P$, $pay_p(\mathbf{x}) < pay_p(\mathbf{y})$.   □

# 3 On the Cost of Bounding the Uncertainty

In this section, we show that even very simple requirements on the outcome of a game dramatically affect the complexity of problems related to the existence and to the computation of Nash equilibria.

## 3.1 Basic Requirements

Our first result shows that the existence of Nash equilibria is no longer guaranteed even in presence of one simple constraint on actions. Furthermore it is unlikely to compute such an equilibrium in polynomial time, as just deciding its existence is an NP-hard problem. The result is proven by means of a construction relating games and Boolean formulas. It is worthwhile noting that in this construction we do not assume any fixed bound on the number of players, while the number of actions may be also very small, even a constant. For a different proof designed for two-player strategic games with an unbounded number of actions, we refer the reader to Conitzer and Sandholm (2003). The comparison of these approaches also helps in understanding the differences and the peculiarities of the two settings.

Recall that deciding whether a Boolean formula in conjunctive normal form $\Phi = c_1 \wedge \ldots \wedge c_m$ over variables $X_1, \ldots, X_n$ is satisfiable, i.e., deciding whether there exists truth assignments to the variables making each clause $c_j$ true is an NP-hard problem, even if each clause contains at most three distinct (possibly negated) variables.

W.l.o.g, assume that the number of clauses is such that there exists a positive integer $l$ with $m = 2^l$. In fact, for the latter assumption, if $m$ is such that $2^{l-1} < m < 2^l$, then we can construct in polynomial time a new Boolean formula $\Phi'$ by adding $2^l - m$ new clauses, each one containing a fresh variable. Obviously, these clauses are trivially satisfiable, and hence $\Phi$ and $\Phi'$ are equivalent.

We define a game $\mathcal{G}(\Phi)$ such that: The players belong to five pairwise disjoint sets $P_v$, $P_{v'}$, $P_{v''}$, $P_c$, and $P_t$ plus one distinguished player $E$. The set $P_v$ (resp. $P_{v'}$, $P_{v''}$) contains exactly one player, say $x_i$ (resp. $x'_i$, $x''_i$) for each variable $X_i$ in $\Phi$, and players in $P_c$ are in a one-to-one correspondence with the clauses. For each player $x_i \in P_v$, her set of neighbors $Neigh(x_i)$ consists of the players $x'_i$ and $x''_i$, for which $Neigh(x'_i) = \{x''_i\}$ and $Neigh(x''_i) = \{x'_i\}$ hold. Each variable $x_i$, occurring in a clause $c_j$ is in the set $Neigh(c_j)$, and no other players are in such set. Players in $P_t$ and player $E$ are such that the subgraph of G($\mathcal{G}(\Phi)$) induced over the nodes in $P_c \cup P_t \cup E$, is a complete binary tree rooted in $E$, having as leaves the players in $P_c$. For each player $t$ in $P_t \cup \{E\}$, $Neigh(t)$ consists of the set of the two players that are children of $t$ in tree induced over G($\mathcal{G}(\Phi)$). Notice that, by construction, $E$ is not a neighbor of any other player, and hence her choices cannot affect the payoffs of the other players in the game.

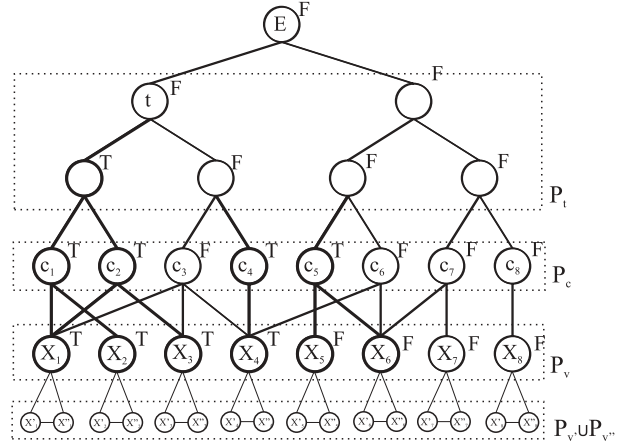

Figure 2: Dependency graph for $\mathcal{G}(\bar{\Phi})$.

Figure 2 shows the dependency graph G($\mathcal{G}(\bar{\Phi})$) of the game $\mathcal{G}(\bar{\Phi})$ associated to the formula $\bar{\Phi} = \exists X_1 \ldots X_8 (X_1 \vee X_2) \wedge (X_1 \vee X_3) \wedge (X_1 \vee \neg X_4) \wedge (X_4) \wedge (\neg X_5 \vee \neg X_6) \wedge (X_4 \vee X_6) \wedge (X_6 \vee X_7) \wedge (X_8)$.

Let $\{T, F\}$ be the set of possible actions, for each player in $\mathcal{G}(\Phi)$, and let $\mathbf{x}$ be a global pure strategy. Utility functions are defined as follows. For each variable $X_i$, the utility functions of the players $x_i$, $x'_i$ and $x''_i$ are shown in tabular form in Figure 3. Intuitively, $x_i$ playing $T$ (resp. $F$) means that we are assuming variable $X_i$ be true (resp. $F$).

Players in $P_v$ assign truth values to the variables in $\phi$, and each player $c \in P_c$ should correctly "evaluate" the corresponding clause in $\Phi$. Indeed, her utility function $u_c$ is such that **(C-i)** $pay_c(\mathbf{x}) = 1$, if the players in $Neigh(c)$ make the corresponding clause true (resp. false), and $c$ plays $T$ (resp. $F$); **(C-ii)** $pay_c(\mathbf{x}) = 0$, in all the other cases.

Each player $t \in P_t$ acts as an AND-gate on the input of her neighbors. Her utility function $u_t$ is such that **(T-i)** $pay_t(\mathbf{x}) = 1$, if either $t$ plays $T$ and all the players in $Neigh(t)$ play $T$, or $t$ plays $F$ and there exists a player in $Neigh(t)$ playing $F$; **(T-ii)** $pay_t(\mathbf{x}) = 0$, in all the other cases.

Finally, player $E$ is responsible for the evaluation of $\Phi$ on the truth assignment induced by the players in $P_v$, and she gets a higher payoff in the case $\Phi$ is satisfied. More precisely, her utility function is such that **(E-i)** $pay_E(\mathbf{x}) = 2$, if $t$ plays $T$ and all the players in $Neigh(E)$ play $T$; **(E-ii)** $pay_E(\mathbf{x}) = 1$, if $t$ plays $T$ and there is a player in $Neigh(E)$ playing $F$; **(E-iii)** $pay_E(\mathbf{x}) = 0$, in all the other cases.

| $x_i'$ \ $x_i''$ | T | F |
|---|---|---|
| T | 1 | 0 |
| F | 0 | 1 |

| $x_i''$ \ $x_i'$ | T | F |
|---|---|---|
| T | 1 | 0 |
| F | 0 | 1 |

| $x_i$ \ $x_i' x_i''$ | TT | TF | FT | FF |
|---|---|---|---|---|
| T | -2 | 0 | 0 | 1 |
| F | 2 | 0 | 0 | -1 |

Figure 3: Utility functions for the game $\mathcal{G}(\Phi)$.

Let $\mathbf{x}$ be a global strategy for $\mathcal{G}(\Phi)$ such that each player $x_i \in P_v$ plays either $T$ or $F$ with probability 1 in $\mathbf{x}$. Note that the choices for $\mathbf{x}$ univocally identify a truth-value assignment for $\Phi$, that we denote by $\sigma^{\mathbf{x}}$. The following result states a useful relationship between satisfying truth-value assignments for $\Phi$ and Nash equilibria of $\mathcal{G}(\Phi)$.

**Lemma 3.1** *Let $\sigma$ be a truth-value assignment for the variables in $\Phi$. Then, $\sigma$ is satisfying $\Leftrightarrow$ if there exists a Nash equilibrium $\mathbf{x}$ such that $E$ plays $T$ with probability 1 in $\mathbf{x}$.*

**Proof.** We first show a number of properties of Nash equilibria of $\mathcal{G}(\Phi)$. The first one regards the gadget shown in Figure 3. Players $x_i'$ and $x_i''$ are designed in such a way that player $x_i$ will eventually choose a pure strategy in any Nash equilibrium. This is a very important feature of game $\mathcal{G}(\Phi)$, since it allows us to reason mostly about boolean values rather than about probabilities.

**Property A:** *If a global strategy $\mathbf{x}$ is a Nash equilibrium for $\mathcal{G}(\Phi)$ then, for each player $x_i \in P_v$, $x_i$ plays either $T$ or $F$ with probability 1 in $\mathbf{x}$.*

In order to prove the claim, let us first calculate the expected payoffs of the players $x_i'$ and $x_i''$ in any global strategy $\mathbf{x}$: $pay_{x_i'}(\mathbf{x}) = x_{i_T}' x_{i_T}'' + x_{i_F}' x_{i_F}'' = x_{i_T}' x_{i_T}'' + (1-x_{i_T}')(1-x_{i_T}'')$; $pay_{x_i'}(\mathbf{x}) = x_{i_T}' x_{i_T}'' + x_{i_F}' x_{i_F}'' = x_{i_T}' x_{i_T}'' + (1-x_{i_T}')(1-x_{i_T}'')$. Then, it can be easily seen that the only possible strategies leading to an equilibrium for such players are: $s_1$, in which they both play $T$ with probability 0 getting payoff 1, $s_2$ in which they both play $T$ with probability 1 getting payoff 1, and $s_3$ in which they both play $T$ with probability $\frac{1}{2}$ getting payoff $\frac{1}{2}$. Similarly, we can calculate the expected payoff of player $x_i$, whose strategy, in fact, depends only on the strategy of $x_i'$ and $x_i''$: $pay_{x_i}(\mathbf{x}) = (1-2x_{i_T})(2x_{i_T}' x_{i_T}'' - (1-x_{i_T}')(1-x_{i_T}''))$. Letting $\alpha = 2x_{i_T}' x_{i_T}'' - (1-x_{i_T}')(1-x_{i_T}'')$ we have to distinguish three cases, depending on the strategy of players $x_i'$ and $x_i''$. **1.** Players $x_i'$ and $x_i''$ choose $s_1$: Then $\alpha = -1$, and player $x_i$ finds convenient to play $T$ with probability 1, getting payoff 1; **2.** Players $x_i'$ and $x_i''$ choose $s_2$: Then $\alpha = 2$, and player $x_i$ finds convenient to play $T$ with probability 0, getting payoff 2; **3.** Players $x_i'$ and $x_i''$ choose $s_3$: Then $\alpha = \frac{1}{2}$, and player $x_i$ finds convenient to play $T$ with probability 0, getting payoff $\frac{1}{2}$. It follows that $\mathbf{x}$ is a Nash equilibrium $\Rightarrow$ each player $x_i \in P_v$ plays either $T$ (case 1 above) or $F$ (case 2 and case 3 above) in $\mathbf{x}$.

Intuitively, property A above tells us that variable players choose pure strategies in any Nash equilibrium $\mathbf{x}$ of $\mathcal{G}(\Phi)$. We show that this correspondence is in fact one-to-one.

**Property B:** *Let $\sigma$ be a truth-value assignment for $\Phi$. Then, there exists a Nash equilibrium $\mathbf{x}$ such that $\sigma^{\mathbf{x}} = \sigma$.*

Given the truth assignment $\sigma$, we consider the global strategy $\mathbf{x}$ for $\mathcal{G}(\Phi)$ where each player in $P_v$ chooses its individual strategy according to $\sigma^{\mathbf{x}} = \sigma$, each player in $P_c$ applies the rules **(C-i)**, i.e., she correctly evaluates the clause, and all players in $P_t \cup \{E\}$, according to the rules **(T-i)**, **(E-i)** and **(E-ii)** act as AND-gates on the inputs of their children. Figure 2 evidences the strategy for $\mathcal{G}(\Phi)$ associated to the truth assignment $X_1 = X_2 = X_3 = X_4 = T$ and $X_5 = X_6 = X_7 = X_8 = F$. It can be easily seen that no player gets an incentive in deviating from the strategy $\mathbf{x}$, which is indeed a Nash equilibrium.

We are now ready to conclude the proof.

($\Rightarrow$) Let $\sigma$ be a satisfying assignment for $\Phi$, and let $\mathbf{x}$ be a Nash equilibrium such that $\sigma = \sigma^{\mathbf{x}}$. Note that $\mathbf{x}$ must exist because of property B above and is such that players in $P_v$ play pure strategies, because of property A. Therefore, at this equilibrium, all the players in $P_c$ have to correctly evaluate the corresponding clauses, thereby getting payoff 1, all the players in $P_t$ have to act correctly as AND-gates, and, hence, $E$ must play $T$ as well, getting payoff 2.

($\Leftarrow$) We show that, for any Nash equilibrium $\mathbf{x}$ for $\mathcal{G}(\Phi)$ where $E$ plays $T$, $\sigma^{\mathbf{x}}$ is a satisfying truth-value assignment for $\Phi$. From property A above, each player in $P_v$ plays in a deterministic way in $\mathbf{x}$. Observe now that each player $t \in P_t \cup \{E\}$ correctly acts as an AND-gate in $\mathbf{x}$. For the sake of contradiction, assume that there exists a player $t \in P_t \cup \{E\}$ that plays $T$ but there exists a player in $Neigh(t)$ playing $F$ in $\mathbf{x}$. Then, $t$ gets payoff 0 and gets an incentive to deviate by playing $F$, which is impossible since $\mathbf{x}$ is an equilibrium. It follows that $E$ plays $T$ if and only if all the players in $P_c$ play $T$ with probability 1. Similarly, we have to assume that players in $P_c$ are correctly evaluating the clauses, and hence that the assignment $\sigma^{\mathbf{x}}$ satisfies all the clauses. □

**Theorem 3.2** *Deciding whether a graphical game $\mathcal{G}$ has constrained Nash equilibria is NP-hard, even if*

there is only one constraint (either on an action or on some player's payoff), and if each player is allowed to play two actions at most and has three neighbors at most.

**Proof.** The reduction is from 3SAT. Let $\Phi$ be a Boolean formula in conjunctive normal form, and $\mathcal{G}(\Phi)$ its associated game. We next show that, adding one constraint of either kind, deciding the existence of a constrained Nash equilibria for $\mathcal{G}(\Phi)$ amounts to deciding satisfiability of $\Phi$:

**Constraint on one action.** We add the constraint $[p_T(E) = 1]$ to game $\mathcal{G}(\Phi)$, which means that player $E$ is forced to choose $T$ with probability 1. Thus, after Lemma 3.1, it follows that $\Phi$ is satisfiable $\Leftrightarrow \mathcal{G}(\Phi)$ admits a constrained Nash equilibrium. We remark that the above proof may be easily modified in order to work with any constraint on actions.

**Constraint on the payoff of one player.** To show that hardness holds even if the one constraint above is replaced by a constraint on payoffs, modify $\mathcal{G}(\Phi)$ so that player $E$ gets payoff $\alpha$ in rule **(E-i)**, $\frac{\alpha}{2}$ in rule **(E-ii)**, and $-\alpha$ in rule **(E-iii)**. Then, $\Phi$ is satisfiable $\Leftrightarrow \mathcal{G}(\Phi)$ admits a Nash equilibrium satisfying the constraint $[pay_E(\mathbf{x}) = \alpha]$. □

## 3.2 Requirements on the Actions of Sets of Players

In this section, we deal with requirements on the actions of sets of players. This setting allows us to model those situations where we know the existence of some Nash equilibrium, but we are unsatisfied with it and we would like to get something different, or to know that no further equilibrium exists. For instance, we may desire that at least one player (not known in advance) deviates from a given unsatisfactory profile, or does not choose her strategy just tossing a coin in a random way. For any game $\mathcal{G}$ and set of players $P$, consider the following problems:

ANOTHER-NASH: Given a Nash equilibrium $\mathbf{x}$ for $\mathcal{G}$, decide whether there exists a Nash equilibrium $\mathbf{y}$ such that, for some player in $P$, her individual strategy in $\mathbf{y}$ is different from her individual strategy in $\mathbf{x}$.

NON-RANDOM: Decide whether there exists a Nash equilibrium $\mathbf{x}$ for $\mathcal{G}$ that satisfies the non-random constraint for $P$. That is, the individual strategy in $\mathbf{x}$ of at least one player $i \in P$ has the form $(pr_a(i), i)$, with $pr_a(i) \neq 1/|Act(i)|$, for some action $a \in Act(i)$.

**Theorem 3.3** ANOTHER-NASH *and* NON-RANDOM *are* NP-*hard.*

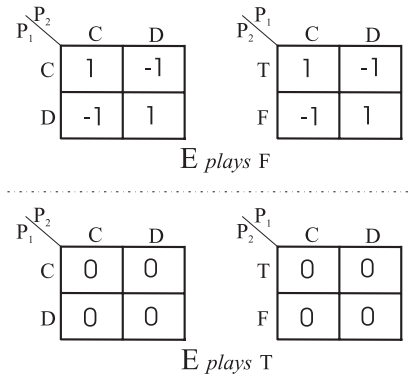

Figure 4: Utility functions for players $P_1$ and $P_2$ in the proof of Theorem 3.3.

**Proof.** Let $\Phi$ be a Boolean formula in conjunctive normal form. We modify $\mathcal{G}(\Phi)$ built in Section 3.1 by adding two new players, say $P_1$ and $P_2$, that do not influence $E$ and such that $Neigh(P_1) = \{P_2, E\}$ and $Neigh(P_2) = \{P_1, E\}$. For a better understanding of this construction, Figure 4 shows the projections of the utility functions of these players, for the two available $E$'s choices. Note that, if $E$ plays $T$, $P_1$ and $P_2$ exhibit a "don't care" behavior, while the other case is somehow conflicting. Let $\mathcal{G}'$ be this modified game. Then, given a global strategy $\mathbf{x}$ for players in $\mathcal{G}'$, the expected payoffs are as follows: $pay_{P_1}(\mathbf{x}) = (1 - E_T) \times (2P_{1_T} - 1)(2P_{2_T} - 1)$; $pay_{P_2}(\mathbf{x}) = -(1 - E_T) \times (2P_{1_T} - 1)(2P_{2_T} - 1)$.

Since players $P_1$ and $P_2$ do not affect the payoffs of any other player in $\mathcal{G}'$ (and in particular of $E$), from Lemma 3.1, we know that $E_T$ can assume only two values at the equilibrium: **1.** If $E_T = 1$, then players $P_1$ and $P_2$ might play any arbitrary strategy, and they get payoff $pay_{P_1}(\mathbf{x}) = pay_{P_2}(\mathbf{x}) = 0$. **2.** if $E_T = 0$, then $P_1$ and $P_2$ are involved in a game which has no pure Nash equilibria. It is easy to see that the only equilibrium is for $P_{1_T} = P_{2_T} = \frac{1}{2}$, in which $pay_{P_1}(\mathbf{x}) = pay_{P_2}(\mathbf{x}) = 0$.

Then, we may prove our two statements as follows.

ANOTHER-NASH. Let $\sigma$ be a truth assignment for $\Phi$. W.l.o.g. assume $\sigma$ is *not* satisfying. Let $\mathbf{x}$ be a Nash equilibria for $\mathcal{G}(\Phi)$ such that $\sigma = \sigma^{\mathbf{x}}$ — notice that such an equilibrium can be constructed as shown in the proof of Property B in Lemma 3.1. Let $\mathbf{x}'$ be the global strategy for $\mathcal{G}'$ obtained by extending $\mathbf{x}$ in a way that players $P_1$ and $P_2$ play $T$ with probability $\frac{1}{2}$. Clearly, $\mathbf{x}'$ is a Nash equilibrium for $\mathcal{G}'$. Moreover, $P_1$ and $P_2$ can change their strategy only if $E$ may play $T$ with probability 1, and hence, after Lemma 3.1, if there exists a satisfying assignment for $\Phi$.

NON-RANDOM. Just observe that $\mathcal{G}'$ has a Nash equilibrium satisfying the non-random constraint for $\{P_1, P_2\}$ (i.e., such that they play $T$ with proba-

bility different from $\frac{1}{2}$) if and only if $\Phi$ is satisfiable. □

By using the arguments in the proof of NON-RANDOM, we can modify the gadget shown in Figure 4 (for the case in which $E$ plays $F$) to easily prove further interesting results. For instance, we may easily prove that it is hard to decide the existence of Nash equilibria which does not belong to a given set $\mathbf{X}$: just design the game in such a way that players $P_1$ and $P_2$ have the set of strategies $\mathbf{X}$ as their possible outcomes.

Therefore, the problem of computing any Nash equilibrium appears to be intrinsically difficult as soon as some kind of strategy is undesirable, and is not considered an acceptable outcome of the game.

## 4  Pareto and Strong Nash Equilibria

In this section, we turn to the study of further kinds of requirements on the payoffs of sets of players. In particular, we first face the problem of checking whether a given profile is a Pareto or a strong Nash equilibrium.

**Theorem 4.1** *Let $\mathcal{G}$ be a graphical game, and let $\mathbf{x}$ be a profile. Then, checking whether $\mathbf{x}$ is a Pareto or a strong Nash equilibrium is* co-NP-*hard, even if each player has only two available actions and at most four neighbors, $\mathbf{x}$ is a Nash equilibrium for $\mathcal{G}$, and there are no further constraints on $\mathcal{G}$.*

**Proof [Sketch].** Let $\Phi$ be a Boolean formula in conjunctive normal form and recall that deciding whether $\Phi$ is not satisfiable is a co-NP-hard problem. Then, consider the game $\mathcal{G}(\Phi) = \langle P, Neigh, Act, U \rangle$ built in Section 3.1, and for any positive value $\gamma$, let $\mathcal{G}'_\gamma = \langle P, Neigh', Act, U' \rangle$ be a new game such that: $Neigh'(p) = Neigh(p) \cup \{E\}$, for each player $p \in P - \{E\}$, and $U' = \{u_E\} \cup \{u'_p \mid p \neq E, u_p \in U\}$ where (i) $u'_p(\mathbf{x}') = \gamma u_p(\mathbf{x})$, if $E$ plays $T$ in $\mathbf{x}'$, and (ii) $u'_p(\mathbf{x}') = u_p(\mathbf{x})$ if $E$ plays $F$ in $\mathbf{x}'$, with $\mathbf{x}$ being the projection of the strategy $\mathbf{x}'$ over the players in $\{p\} \cup Neigh(p)$. Note that $\mathcal{G}'_\gamma$ is obtained in such way that each player depends on $E$. We show that, for any value of $\gamma > 1$, the Nash equilibria of $\mathcal{G}(\Phi)$ are preserved in $\mathcal{G}'_\gamma$.

**Property C:** *Let $\mathbf{x}$ be a global strategy for $\mathcal{G}(\Phi)$ and $\mathcal{G}'_\gamma$ the game constructed from $\mathcal{G}(\Phi)$, for some $\gamma > 1$. Then, $\mathbf{x}$ is a Nash equilibrium for $\mathcal{G}(\Phi) \Leftrightarrow \mathbf{x}$ is a Nash equilibrium for $\mathcal{G}'_\gamma$.*

Let us compute the expected payoff of each player $p$ in $\mathcal{G}'_\gamma$, denoted by $pay'_p$. By exploiting the definition of the utility functions in $U'$, we easily derive $pay'_p(\mathbf{x}) = E_T \gamma pay_p(\mathbf{x}) + (1 - E_T) pay_p(\mathbf{x}) = pay_p(\mathbf{x})(1 + (\gamma - 1) E_T)$. It follows that, for $\gamma > 1$, the actual value of $E_T$ has no influence on the strategy of each player, and hence Nash equilibria are preserved in the modified game.

In the following, we consider for simplicity the case $\gamma = 2$, that leads to doubling the payoffs of each player whenever $E$ plays $T$. Consider a truth-value assignment $\sigma$ which is not satisfying, and let $\mathbf{x}$ be the Nash equilibrium associated to $\sigma$, constructed by exploiting Property B in Lemma 3.1. We first show that $\mathbf{x}$ *is a Pareto Nash equilibrium $\Leftrightarrow$ the formula is not satisfiable.*

($\Rightarrow$) Let $\mathbf{x}$ be a Pareto Nash equilibrium and assume, for the sake of contradiction, that $\Phi$ is satisfiable. Then, take one satisfying assignment, say $\sigma^*$, and consider the equilibrium $\mathbf{x}^*$ that is associated to $\sigma^*$ according to Lemma 3.1 and preserved by Property C above. Since $E$ plays $T$ in $\mathbf{x}^*$ with probability 1, each player in $\mathcal{G}'_2$ but $E$ will double her payoff in $\mathbf{x}^*$ — see the form of the payoffs in Property C. Moreover, player $E$ gets also a better payoff since the formula is satisfied in $\mathbf{x}^*$ and since she can apply (E-i) rather than (E-ii). It follows that each player gets a higher payoff if all of them jointly deviate from $\mathbf{x}$ to $\mathbf{x}^*$. Contradiction.

($\Leftarrow$) Assume that $\Phi$ is not satisfiable and, for the sake of contradiction, that $\mathbf{x}$ is not Pareto. Then, let $\mathbf{x}^*$ be the Nash equilibrium for which all the players gets an incentive to deviate from $\mathbf{x}$. Since $\mathbf{x}$ is not Pareto, we have to assume that $E$ gets payoff 1 in $\mathbf{x}$ (otherwise she could not increase this payoff in $\mathbf{x}^*$) and payoff 2 in $\mathbf{x}^*$. Then, due to Lemma 3.1, $\sigma^{\mathbf{x}^*}$ is a satisfying assignment. Contradiction.

We conclude the proof by showing that $\mathbf{x}$ *is a strong Nash equilibrium $\Leftrightarrow$ the formula is not satisfiable.*

($\Rightarrow$) Let $\mathbf{x}$ be a strong Nash equilibrium and assume, that $\Phi$ is satisfiable. We have just shown that in this scenario there exists an equilibrium $\mathbf{x}^*$ in which each player gets a strictly higher payoff than in $\mathbf{x}$. We thus have a contradiction, since this equilibrium testifies there is an incentive to deviate for the coalition comprising all players.

($\Leftarrow$) Assume that $\Phi$ is not satisfiable and, for the sake of contradiction, that $\mathbf{x}$ is not strong. Let $K$ be a coalition such that each player $p \in K$ may increase her payoff by choosing an action that differs from the one played in $\mathbf{x}$. We can show that $K$ must comprise a number of players in $P_v$, $P_{v'}$ and $P_{v''}$, $P_t$ plus $E$. In particular, since $E$ can increase her payoff in the case she is evaluating a satisfying assignment, it must be the case that players in $P_v \cap K$ are such that the formula is now satisfied. Contradiction. □

As far as the existence problem is concerned, recall that a Pareto equilibrium always exists in the mixed strategies setting, though checking whether a profile has this property is an intractable problem. The fol-lowing result states that the situation is rather different for strong Nash equilibria. Indeed, it turns out that this kind of additional stability requirement makes the problem hard for the second level of the polynomial hierarchy. For space reasons we omit here the proof of this theorem, which is included in the full version of this paper, available from the authors.

**Theorem 4.2** *Deciding whether a graphical game has a strong Nash equilibrium is $\Sigma_2^P$-hard, even if each player has only two available actions and at most five neighbors, and there are no further constraints on $\mathcal{G}$.*

## 5 Conclusions

In this paper, we dealt with mixed Nash equilibria in graphical games satisfying additional requirements, ranging from basic constraints on single players (both on actions and payoffs) to advanced requirements on sets of players, thus complementing a previous work on constrained *pure* Nash equilibria by Greco and Scarcello (2004). Our work also answers some research questions posed by Conitzer and Sandholm (2003).

It turned out that bounding the uncertainty of game outcomes immediately unsettles our only certainty, as the existence of a (feasible) Nash equilibrium is no longer guaranteed, and its computation is unlikely to be tractable. Our results help in shedding some light on the nature of Nash equilibria and the complexity of dealing with them. For instance, even looking for strategies having the simple requirement to be not fully random is an NP-hard problem.

It is worthwhile noting that we provide no membership result in this paper. The problem here is that some equilibria may not be finitely representable, e.g. the 3-player game proposed by Nash (1950), which has no equilibrium with all rational probabilities. Therefore, membership could be faced with different approaches, possibly based on suitable approximations of equilibria, that we leave as an interesting issue for further research.

Another avenue of future work is to study whether our results may translate to the setting of two-player strategic games, possibly with an unbounded number of actions. For instance, it would be interesting to know whether our hardness results for Pareto and strong Nash equilibria hold in this setting, too.